# Shadows and accretion disk images of charged rotating black hole in modified gravity theory

He-Bin Zheng[a], Meng-Qi Wu[b], Guo-Ping Li[c], Qing-Quan Jiang[d]

School of Physics and Astronomy, China West Normal University, Nanchong 637000, China

**Abstract** In this paper, we study the shadows and images of the accretion disk of Kerr–Newman (KN) black hole (BH) in modified gravity (MOG) theory by using the backward ray-tracing method. And, the influence of spin parameter ($a$), charge ($Q$), and MOG parameter ($\alpha$) on the observed features of BHs are carefully addressed. Interestingly, as $\alpha$ increases, the flat edge of the BH's shadow gradually becomes more rounded, the size of shadow enlarges, and the deviation rate ($\delta s$) correspondingly decreases. By tracing the photon around BH, we observe that the trajectory of photon exhibits distortion behavior, i.e., the formation of two "tails" near the Einstein ring, which elongate as $a$ increases. For the accretion disk, it shows that the inner shadow expands with $\alpha$, while decreases with $Q$. The increase of $\alpha$ exhibits an increasing effect on redshift. At the same parameter level, $\alpha$ has a more obvious effect on inner shadow and image of BH by comparing with that of $Q$. Our study implies that both $\alpha$ and $Q$ have relatively significant effects on the image of the KN-MOG BH with the thin disk accretion, but the influence of $\alpha$ is much greater. So, this indicates that $\alpha$ plays a dominant role in this spacetime.

## 1 Introduction

Black hole (BH) is a type of dense celestial object predicted by general relativity, scientists have primarily focused on their theoretical study for decades. But in recent years, with advancements in technology, the Laser Interferometer Gravitation Wave Observatory (LIGO) has detected gravitational wave signals generated by the merger of binary BHs in 2015 [1]. Later on, the Event Horizon Telescope (EHT) also captured the first image of a supermassive BH's shadow in 2019 [2–7]. This groundbreaking discovery holds significant implications for both observational astronomy and theoretical physics. The BH's image reveals a bright ring encircling a dark region, identified as the photon ring and the BH's shadow, respectively [8]. It is well-known that BH's shadow can offer some deep insights into BH's properties. The shadow formed by the deflection of light traveling from the distant region around the BH, which is the result of the BH's intense gravitational field. As more and more researchers focus on optical observations of BHs, the methods for studying BH's shadows are becoming richer and more diverse. In fact, long before the EHT captured images of BHs, many scientists had already conducted theoretical simulations of light trajectories and BH's shadow images. In 1966, Synge studied the photon trajectory of a Schwarzschild BH [9]. In 1979, Luminet first predicted through computer simulations that the gravitational field around a black hole would bend light, creating a gravitational lensing effect [10]. When one studied the strong gravitational lensing caused by a Schwarzschild BH, Virbhadra discovered relativistic images on both sides of the optic axis [11]. Then, the concept of photon sphere in Schwarzschild spacetime is extended to the concept of photon surface in arbitrary spacetimes [12]. And, Virbhadra et al. studied the variations in tangential, radial, and total magnifications of the primary, secondary, and relativistic images with respect to changes in angular source position and the ratio of lens-source to observer-source distance using certain distortion parameters [13,14]. Also, they calculated the compactness of the lens through the first-order relativistic image [15]. Obviously, the bending of light further leads to the formation of the shadow of BH. For the shadow, Takahashi used BH's shadows to constrain the electric charge of a BH [16]. And, the shadow can provide some strong evidences for distinguishing BHs in various gravita-tional theories or matter fields with astronomical observations. For instance, Zhong studied the shadow of a Kerr BH in the presence of a uniform magnetic field [17]; Wei studied the properties of Gauss-Bonner gravity using the shadows of four-dimensional rotating BHs [18]; Wang and Wei demonstrated that BH's shadows can be used to test the Lorentz symmetry [19]; Yuan combined the first-order QED effect with the numerical inverse ray tracing method to study the shadow of a magnetic charged KN BH [20]. Using the Hamiltonian constraint approach, Liu conducted an in-depth analysis of the light rings and shadows of two static BHs that preserve general covariance, employing topological methods and the backward ray-tracing technique, and the study confirms that the light rings of both types of static BHs are classified as standard and unstable [21]. And, there are still many highly valuable articles in the field of gravitationally lensed images and BH's shadows [22–28]. In addition, through the BH's image provided by the EHT, it is found that the BH's shadow is related to the BH's mass, spin parameters, charge parameters, accretion disk and modified gravitational parameters [8,29–36].

[a] e-mail: zhenghb3060@163.com
[b] e-mail: 15282598923@163.com
[c] e-mail: gpliphys@yeah.net (corresponding author)
[d] e-mail: qqjiangphys@yeah.net

It is widely believed that BHs in the universe are always surrounded by the accretion flows due to strong gravity. Therefore, some researchers have studied the various energy fluxes and the emission spectrum of BHs, and regarded their accretion disk as a perfect black body radiator [37–39]. And some researchers have studied Schwarzschild BHs by considering spherically symmetric accretion [40]. It is gradually found that when the BH is surrounded by an optically and geometrically thin accretion disk, it can be observed that the BH's image is composed of bright and dark regions, including direct emission rings, lens rings and photon rings [41]. Selecting different emission functions can also reveal interesting features in the shadows and rings. For thin accretion disk, the contribution of the brighter lens ring and the darker photon ring surrounding the BH's shadow to the total observed intensity is always quite small, so their contributions to the total observed intensity can be ignored [42]. These features are significant for our study of the images and properties of BHs surrounded by thin accretion disks. Meanwhile other studies show that thin accretion disks have significant effects on BH's shadow [31,43–49]. Through the tireless efforts of researchers, significant progress has been made in the theoretical study of BH's shadows, accretion disks and BH's images, making important contributions to astronomical observations [42,49–81].

On the other hand, Moffat introduced a modified theory of gravity called the scalar-tensor-vector theory of gravity [82], also known as the MOG theory. MOG theory has successfully explained solar system observations, galaxy rotation curves and galaxy cluster dynamics [83–87]. Also, it gives data of matter power spectrum and cosmic microwave background sound power spectrum [88]. In the study of galactic rotation curves, people can choose modified gravity theory rather than general relativity [89–91]. In general, the two theories differ in key respects. Many studies have highlighted the difference between MOG and general relativity [92–102]. In 2014, Moffat et al. proposed static BH's solutions and axisymmetric BH's solutions in MOG theory [103]. Then, one has in MOG theories studied the thermodynamic properties, quasinormal scale, shadows and accretion disks of BHs [104–107]. Although the MOG theory has made great progress in constructing BH's solutions and studying the physical properties of BHs, the properties of rotating charged BH's solutions in MOG theory remain unknown. The most general axially symmetric BH solution to Einstein's equations is the KN solution, which includes spin ($a$), charge ($Q$), and angular momentum. This solution in the MOG theory has already been confirmed by researchers [92,108]. Therefore, it is of great significance to study BH's shadows and the images of accretion disks in MOG theory. In particular, the shadow of the rotating BH in MOG theory and its observed image under a thin disk are still unclear. In view of this, the aim of this paper is to study the shadows and accretion disk images of rotating charged MOG BHs, with the goal of providing some astronomical observational insights into MOG theory.

The structure of this paper is as follows: in Sect. 2, we studied the spacetime properties of the KN BH in the MOG theory and analyzed its shadow images. Section 3, is to introduce the detail of the thin disk model. In Sect. 4, we presented the images of the thin disk accretion, as well as its redshift and intensity distribution, and compared the effects of the MOG parameter on the observed images with that of charge. Finally, Sect. 5 ends with a brief summary.

## 2 KN-MOG spacetime and shadow

In 2015, Moffat has successfully employed the scalar-tensor-vector gravity (STVG) or MOG to construct the rotating BH's solution [92]. In Boyer-Lindquist coordinates, the expression of this charged spacetime can be expressed as [108]

$$ds^2 = -\left(\frac{\Delta - a^2 \sin^2\theta}{\Sigma}\right)dt^2 + \frac{\Sigma}{\Delta}dr^2 + \Sigma d\theta^2$$
$$+ \frac{\sin^2\theta}{\Sigma}\left[(a^2+r^2)^2 - a^2\sin^2\theta\Delta\right]d\phi^2$$
$$+ \frac{2a\sin^2\theta}{\Sigma}\left[\Delta - (a^2+r^2)\right]dtd\phi, \quad (2.1)$$

where

$$\Delta = r^2 - 2GMr + a^2 + Q^2 + G_N^2\alpha(1+\alpha)M^2,$$
$$\Sigma = r^2 + a^2\cos^2\theta. \quad (2.2)$$

The parameters $a$ and $M$ denote the spin and mass of BH respectively, while $Q$ represents the charge of the KN-MOG BH. For simplicity, we set $G_N = 1$ and $M = 1$ in the following discussions. The parameter $\alpha = (G - G_N)/G_N$ is the MOG parameter, which denotes the dimensionless measure of the difference between the Newtonian gravitational constant $G_N$ and the additional gravitational constant $G$. When $\alpha = 0$, the KN-MOG BH reduces to the KN BH. When $\alpha = 0$ and $Q = 0$, the KN-MOG BH further reduces to the Kerr BH. To ensure the existence of the event horizon, we define the critical charge, i.e., $Q_e = \sqrt{-a^2 + G_N^2 M^2 (1 + \alpha)}$. The shadow of BH is determined by the motion of massless particles following null geodesics in its spacetime structure. The geodesic equation can be derived by solving the Hamilton-Jacobi equation:

$$\frac{\partial S}{\partial \sigma} = -\frac{1}{2} g^{\mu\nu} \frac{\partial S}{\partial x^\mu} \frac{\partial S}{\partial x^\nu}, \tag{2.3}$$

where $S$ denotes the Jacobi action, which is given by the following expression

$$S = -\frac{1}{2} m^2 \sigma - \mathcal{E} t + \mathcal{L} \phi + S_r(r) + S_\theta(\theta). \tag{2.4}$$

In Eq. (2.4), $m$ represents the mass of test particle, $\mathcal{E}$ denotes the energy, and $\mathcal{L}$ is the angular momentum along the axis of symmetry. We focus on the motion of photon($m = 0$). So, there are two conserved quantities, which are

$$\mathcal{E} = -g^{tt} \dot{t} - g^{t\phi} \dot{\phi} x, \quad \mathcal{L} = g^{t\phi} \dot{t} + g^{\phi\phi} \dot{\phi}, \tag{2.5}$$

the overdot denotes differentiation with respect to the affine parameter $\tau$. By substituting Eqs. (2.1) and (2.2) into Eqs. (2.4) and (2.5), it can be obtained

$$\Sigma^2 \dot{t}^2 = \Sigma^2 \left[ \mathcal{E} + \frac{(a^2 \mathcal{E} - a\mathcal{L} + \mathcal{E} r^2)(a^2 + r^2 - \Delta)}{\Delta \Sigma} \right]^2,$$

$$\Sigma^2 \dot{\phi}^2 = \Sigma^2 \left[ \frac{\Sigma \csc^4 \theta \left[ a[a^2 \mathcal{E} - a\mathcal{L} + \mathcal{E}(r^2 - \Delta)] + \mathcal{L} \Delta \csc^2 \theta \right]}{\Delta \left[ a^2 - (a^2 + r^2) \csc^2 \theta \right]^2} \right]^2,$$

$$\Sigma^2 \dot{r}^2 = \left[ (r^2 + a^2) \mathcal{E} - a\mathcal{L} \right]^2 - \Delta \left[ \kappa + (\mathcal{L} - a\mathcal{E})^2 \right] = R,$$

$$\Sigma^2 \dot{\theta}^2 = \kappa + \cos^2 \theta \left( a^2 \mathcal{E}^2 - \frac{\mathcal{L}^2}{\sin^2 \theta} \right) = \Theta, \tag{2.6}$$

$\kappa$ is the Carter constant. Using these equations, we can determine the photon's trajectory. Next, we relate the energy $\mathcal{E}$, angular momentum $\mathcal{L}$, and Carter's constant $\kappa$ to the impact-parameters near the BH: $\xi = \mathcal{L}/\mathcal{E}$ and $\eta = \kappa/\mathcal{E}^2$. In general, the condition of photon sphere demands $R(r_p) = 0$ and $dR(r_p)/dr_p = 0$, where $r_p$ is the radius of photon sphere. So, we can obtain

$$\xi(r) = \frac{a^2(1 + \alpha + r) + r\mathbb{B}}{\mathbb{A}} \bigg|_{r=r_p},$$

$$\eta(r) = -\frac{r^2 \left[ 4a\alpha \mathbb{A} + 4a^2(Q^2 - r) + \mathbb{B}^2 \right]}{\mathbb{A}^2} \bigg|_{r=r_p}, \tag{2.7}$$

where $\mathbb{A} = a(1 + \alpha - r)$, $\mathbb{B} = \left[ 2\alpha^2 + 2Q^2 + \alpha(2 - 3r) + (-3 + r)r \right]$. First, we assume that the observer located at infinity. Due to the symmetries in the $t$ and $\phi$ directions, we can define a zero-angular-momentum observer (ZAMO) at coordinates $(t_o = 0, r_o, \theta_o, \phi_o = 0)$ with the tetrad reads

$$e_0 = \zeta \partial_t + \gamma \partial_\phi, \quad e_1 = -\frac{\partial_r}{\sqrt{g_{rr}}},$$

$$e_2 = \frac{\partial_\theta}{\sqrt{g_{\theta\theta}}}, \quad e_3 = -\frac{\partial_\phi}{\sqrt{g_{\phi\phi}}}, \tag{2.8}$$

where $\zeta = \sqrt{\frac{-g_{\phi\phi}}{g_{tt}g_{\phi\phi} - g_{t\phi}^2}}$ and $\gamma = -\frac{g_{t\phi}}{g_{\phi\phi}} \sqrt{\frac{-g_{\phi\phi}}{g_{tt}g_{\phi\phi} - g_{t\phi}^2}}$. In our previous work, the model the tetrad and the camera model was provided in [109]. Considering a light ray $\lambda(\tau) = (r(\tau), \theta(\tau), \phi(\tau), t(\tau))$, its tangent vector is given by

$$\dot{\lambda} = \dot{r} \partial_r + \dot{\theta} \partial_\theta + \dot{\phi} \partial_\phi + \dot{t} \partial_t = \left| \overrightarrow{OP} \right| (-\chi e_0 + \cos \alpha e_1 + \sin \alpha \cos \beta e_2 + \sin \alpha \sin \beta e_3). \tag{2.9}$$

Based on Eq. (2.9), $\chi$ is an undetermined frefactor. Notably, the trajectory of the photon is independent of its energy. The photon's energy is introduced into the camera's frame and set to unity, i.e., $E_{camera} = |\overrightarrow{OP}|\chi = 1$. In the ZAMO's frame, the four-momentum of photons can be written as

$$p^0 = \zeta \mathcal{E} - \gamma \mathcal{L}, \quad p^1 = -\frac{1}{\sqrt{g_{rr}}} k_1,$$

$$p^2 = \frac{1}{\sqrt{g_{\theta\theta}}} k_2, \quad p^3 = -\frac{1}{\sqrt{g_{\phi\phi}}} \mathcal{L}, \tag{2.10}$$

$k_\nu$ is the photon's four-momentum before being projected onto the ZAMO's frame. According to the relationship between photon four-momentum and celestial coordinates that discribed in [109], we have $\cos \alpha = p^1/p^0$ and $\tan \beta = p^3/p^2$. So, the expressions for the celestial coordinates in the spacetime (2.1) are provided below

$$\alpha(r_p) = \arccos\left[\frac{\sqrt{\mathbb{C}}\sqrt{(a^2+r^2)^2 - a^2\Delta - \Delta\,\eta(r_p) - 2\,\mathbb{D} + (a^2-\Delta)\xi^2(r_p)}}{\mathbb{C} - \mathbb{D}}\right]\Bigg|_{(r_o,\theta_o)}, \tag{2.11}$$

$$\beta(r_p) = \arctan\left[\frac{\sqrt{\Sigma^2}\,\xi(r_p)}{\sqrt{\sin^2\theta\,\mathbb{C}}\sqrt{a^2\cos^2\theta + \eta(r_p) - \cot^2\theta\,\xi^2(r_p)}}\right]\Bigg|_{(r_o,\theta_o)}, \tag{2.12}$$

where $\mathbb{C} = (a^2+r^2)^2 - a^2\Delta\sin^2\theta$ and $\mathbb{D} = a(a^2+r^2-\Delta)\xi(r_p)$. With the aid of Eqs. (2.11, 2.12), it is easy to find the relation between the celestial coordinates and Cartesian coordinates $(x, y)$ one the screen [75]. So, we have

$$x(r_p) = -2\tan\left[\frac{\alpha(r_p)}{2}\right]\sin\left[\beta(r_p)\right],$$
$$y(r_p) = -2\tan\left[\frac{\alpha(r_p)}{2}\right]\cos\left[\beta(r_p)\right]. \tag{2.13}$$

In Fig. 1, we set the charge to $Q = 0.3Q_e$ and show the shadows of KN-MOG BH for different choices of $a$ and $\alpha$. In Fig. 1a, as $\alpha$ increases, we observe the shadow's size expanding slightly. This expansion indicates that the modified gravity enhances the effects of the BH's gravitational field, causing more light to bend and increasing the size of the shadow. As in Fig. 1b, one can see that the shadow evolves into a "D" shape when $a$ increases. The larger $a$ causes a more noticeable difference in the shape of the shadow. Notably, as $\alpha$ increases, this "D" shape of shadow gradually return to a circular shape. This suggests that modified gravity alters the bending of light around the BH, and the influence becomes more pronounced with increasing $\alpha$.

In Fig. 2, the charge $Q$ decreases the shadow radius. And, it is interesting to find that the smaller of $a$, the greater effect of charge on the shadow. In Fig. 2, increasing $Q$ enhances the BH's electric field, which weakens the degree of light bending around it, leading to a reduction in the size of the BH's shadow. This is because a stronger electric field more effectively repels nearby light, reducing the degree of light bending and causing the shadow region to shrink. In the case of low $a$, the effect of $Q$ on the shadow is relatively more significant, causing the shadow to shrink noticeably. In the case of large $a$, the reduction in the BH's shadow is smaller. Therefore, the size of the shadow of low-spin BHs is more sensitive to changes in charge. Next, to further show the effect of $\alpha$, we study the shadow radius and deviation rate $\delta s$ of shadow which defined in [110]. For different values of $\alpha$, the shadow radius and deviation rate are shown in Table 1.

From the Table 1, as the value of $\alpha$ increases, the shadow radius gradually increases regardless of the spin parameter $a$. This change indicates that the modified gravity theory increases the effect of light bending, which in turn leads to an expansion of the BH's shadow radius. In the case of a small spin parameter (e.g., $a = 0.1$), the shadow radius increases from 0.050695 to 0.079475 (the variation is 0.02878) as $\alpha$ changes from 0 to 0.7, showing a significant increase. In the case of a larger spin parameter (e.g., $a = 0.998$), the shadow radius increases from 0.051475 to 0.080038 (the variation is 0.028563). Although there is an increase, the effect is somewhat smaller compared to BHs with lower spin parameter. For a small spin parameter ($a = 0.1$), the deviation rate gradually decreases as $\alpha$ increases, from 0.001222 to 0.000689 (the variation is 0.000533). For larger spin parameter ($a = 0.998$), the deviation rate decreases significantly as $\alpha$ increases, and the change in deviation rate is more pronounced for a higher spin BH. It decreases from 0.247953 to 0.084967 (the variation is 0.162986). Therefore, the data in the table indicates that the modified gravity parameter $\alpha$ has a significant impact on the BH's shadow, especially for larger spin BHs, where the effect of modified gravity on the shadow's shape is more pronounced, which coincides with the results shown in Fig. 1.

For a celestial light source, we will more easily observe the shadow of BH and the distortion of light around it. When BH illuminated by a celestial light source, Fig. 3 presents images of KN-MOG BH for both $a = 0.1, 0.998$ and $\alpha = 0, 0.1, 0.5$. The color divisions of the four quadrants of the celestial sphere divided into: for $0 \leq \theta < \pi/2$, the region $0 \leq \phi < \pi$ marked as green quadrant, and the red quadrant belongs to $\pi \leq \phi < 2\pi$; for $\pi/2 \leq \theta < \pi$, the blue quadrant is fixed to region $0 \leq \phi < \pi$, and the range $\pi \leq \phi < 2\pi$ is assigned as the yellow quadrant [111]. When the parameter $\alpha$ is fixed at $\alpha = 0.1$, the left side of the shadow will be flattened with the increase of $a$. This change is very small for a smaller $a$, while it will becomes pronounced when $a = 0.998$. So, we can observe that as $\alpha$ increases, the shadow deformation caused by $a$ diminishes. And, when $\alpha$ is very large, the BH's shadow always reverts to a black circular disk even if $a$ is very large. In addition, it is also obvious that the parameter $\alpha$ enlarges the size of the shadow. On the other hand, the two quadrants, the green and the blue regions, exhibit a "tail".

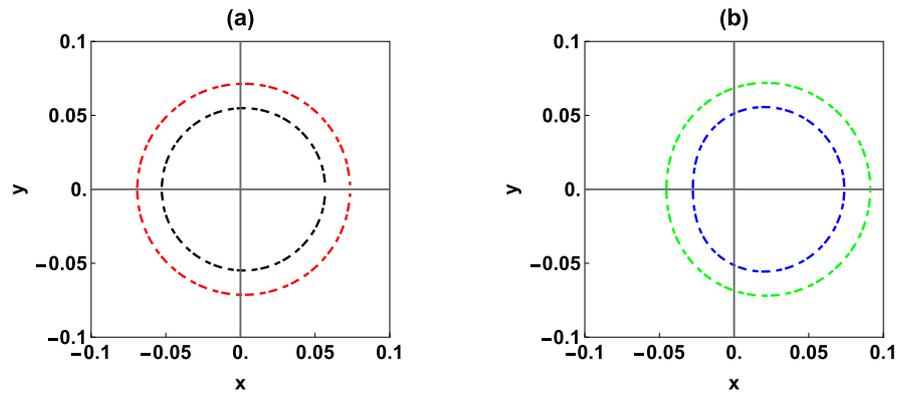

**Fig. 1** The shadow of KN-MOG BH when $Q = 0.3Q_e$. The position of observer is fixed at $(r_o = 100, \theta_o = \pi/2)$. For the left plane, when $a = 0.1$, the black dashed line corresponds to the case ($\alpha = 0.1$), and the red dashed line corresponds to the case ($\alpha = 0.5$). For the right plane, when $a = 0.998$, the blue and green dashed lines correspond to the cases ($\alpha = 0.1$) and ($\alpha = 0.5$), respectively

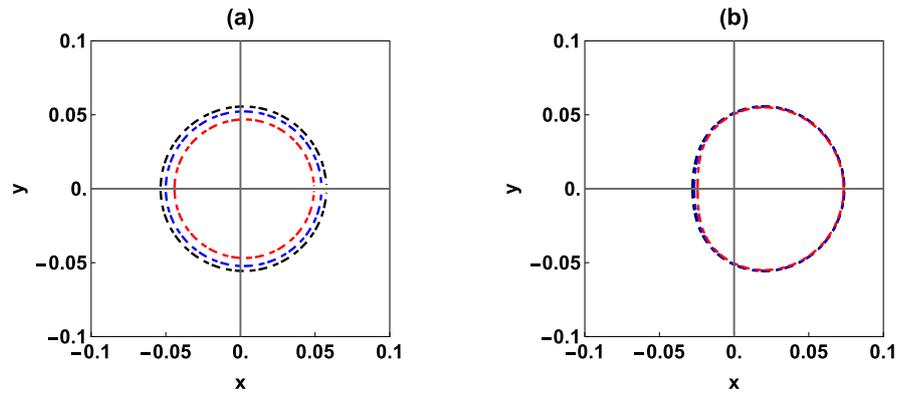

**Fig. 2** The shadow of KN-MOG BH when $\alpha = 0.1$. The position of observer is fixed at $(r_o = 100, \theta_o = \pi/2)$. For the left plane, when $a = 0.1$, the black dashed line corresponds to the case ($Q = 0.1Q_e$), and the blue dashed line represents the case ($Q = 0.6Q_e$), and the red dashed line is the case ($Q = 0.9Q_e$). For the left plane, when $a = 0.998$, the black, blue and red dashed lines correspond to the cases ($Q = 0.1Q_e$), ($Q = 0.6Q_e$) and ($Q = 0.9Q_e$), respectively

This "tail" is obvious related to the spin of BH, which seem to be independent of $\alpha$; as $a$ increases, the "tail" becomes longer. In fact, due to gravitational lensing, there will be an Einstein ring outside of the shadow. The "tail" in the inner quadrant will extend along the inner edge of Einstein ring, while the "tail" in the outer quadrant extend along the outer edge of Einstein ring.

## 3 Thin accretion disk model

In our study, the accretion disk model is represented as a free, electrically neutral plasma orbiting along an equatorial time-like geodesic, and is optically and geometrically thin disk. In our previous work, the schematic diagram of a thin accretion disk was provided [109]. Using $r_{ISCO}$ as the dividing line, the accretion disk can be divided into two regions. The radius of the ISCO can be determined by solving this equation

$$V_{eff}(r, E, L) = 0, \quad \partial_r V_{eff}(r, E, L) = 0,$$
$$\partial_r^2 V_{eff}(r, E, L) = 0, \quad (3.1)$$

where $V_{eff}$ is the effective potential function

$$V_{eff}(r, E, L) = \left(1 + g^{tt}E^2 + g^{\phi\phi}L^2 - 2g^{t\phi}EL\right)\Big|_{\theta=\frac{\pi}{2}}. \quad (3.2)$$

Since the $r_{ISCO}$ divided the accretion disk into inner and outer parts, it is important to note that the motion for the

**Table 1** The shadow radius $Rs$ and the deviation rate $\delta s$ of KN-MOG BH for different values of $a$ and $\alpha$ when $Q = 0.3Q_e$, where the observer locates at $(r_o = 100, \theta_o = \pi/2)$. The first three columns on the left represent the shadow radius of KN-MOG BH, and the last three columns on the right represent KN-MOG BH's circularity deviation

| $\alpha$ | $Rs(a=0.1)$ | $Rs(a=0.6)$ | $Rs(a=0.998)$ | $\delta s(a=0.1)$ | $\delta s(a=0.6)$ | $\delta s(a=0.998)$ |
|---|---|---|---|---|---|---|
| 0 | 0.050695 | 0.050975 | 0.051475 | 0.001222 | 0.049050 | 0.247953 |
| 0.1 | 0.054889 | 0.055153 | 0.055626 | 0.001108 | 0.043955 | 0.176622 |
| 0.3 | 0.063182 | 0.063421 | 0.063848 | 0.000928 | 0.036213 | 0.128211 |
| 0.5 | 0.071371 | 0.071589 | 0.071981 | 0.000793 | 0.030602 | 0.102214 |
| 0.7 | 0.079475 | 0.079676 | 0.080038 | 0.000689 | 0.026348 | 0.084967 |

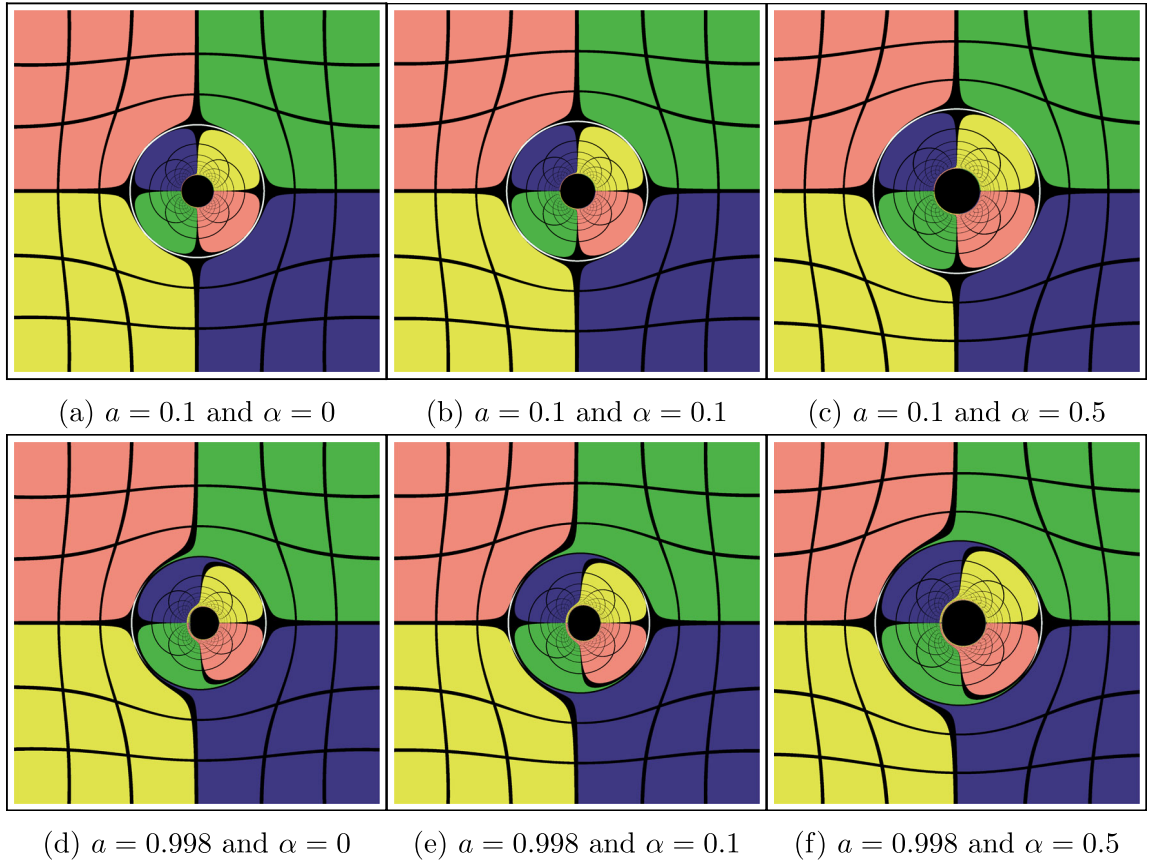

(a) $a = 0.1$ and $\alpha = 0$
(b) $a = 0.1$ and $\alpha = 0.1$
(c) $a = 0.1$ and $\alpha = 0.5$
(d) $a = 0.998$ and $\alpha = 0$
(e) $a = 0.998$ and $\alpha = 0.1$
(f) $a = 0.998$ and $\alpha = 0.5$

**Fig. 3** Images of BH under the celestial light source for $Q = 0.3 Q_e$. The top row and the bottom row correspond to ($a = 0.1$) and ($a = 0.998$), respectively. From left to right, the MOG parameters of BHs are ($\alpha = 0, 0.1, 0.5$), respectively. The position of observer is fixed at ($r_o = 100$, $\theta_o = \pi/2$), and the camera's field of view is $60°$. And the black disk represents the shadow of BH, while the white ring outside of shadow represents the Einstein ring

particles is differ in these two regions. At first, we need to determine the four-velocity $u^a$ of the particles in the accretion disk. Due to the invariance under time translation and rotation, the KN-MOG BH possesses two Killing vectors. Consequently, along a time-like geodesic, the two conserved quantities of a massive neutral particle's four-velocity can be expressed as $E = -u_t$ and $L = u_\phi$. $E$ is the energy per unit mass and $L$ is angular momentum per unit mass. And, the normalization condition $u^a u_a = -1$ should be satisfied. When $r \geq r_{\text{ISCO}}$, the equation of motion of the particle in the radial direction can be obtained by $V_{eff} = \partial_r V_{eff} = 0$. With these two equations, it is easy to find two conserved quantities $E = E_{cir}(r)$ and $L = L_{cir}(r)$ for the circular orbits. For a stable circular orbit, we need to solve the second derivative of the effective potential of the circular orbit $d_2 V_{eff}(r) = \partial_r^2 V_{eff}\big|_{E=E_{\text{cir}}(r), L=L_{\text{cir}}(r)}$. So, the effective potential must satisfy $d_2 V_{eff}(r) \geq 0$. In this sense, we regarded the motion of particles outside of $r_{ISCO}$ is the circular orbital motion. Inside of $r_{ISCO}$, we consider that the motion of particles is the plunge orbit fallen from $r_{ISCO}$ to the event surface. The photon in this state is assumed to have energy $E_{ISCO}$ and angular momentum $L_{ISCO}$. In this consideration, the accretion flows move along time-like circular orbits, the equation of motion can also be rewritten as

$$u^a = u^t_{out}(1, 0, 0, \Omega_n), \quad (3.3)$$

where

$$u^t_{out} = \sqrt{-\frac{1}{g_{\phi\phi}\Omega_n^2 + 2g_{t\phi}\Omega_n + g_{tt}}}\bigg|_{\theta=\frac{\pi}{2}},$$

$$\Omega_n = \frac{-g_{t\phi,r} + \sqrt{g_{t\phi,r}^2 - g_{tt,r}g_{\phi\phi,r}}}{g_{\phi\phi,r}}\bigg|_{\theta=\frac{\pi}{2}}. \quad (3.4)$$

Inside the ISCO, the matter falls from the ISCO to the event horizon with the same conserved quantities as that for the ISCO. The components of four-velocity can be written as

$$u^t_{in} = -g^{tt}E_{ISCO} + g^{t\phi}L_{ISCO}\bigg|_{\theta=\frac{\pi}{2}},$$

$$u_{in}^r = -\sqrt{-\frac{V_{eff}(r, E_{ISCO}, L_{ISCO})}{g_{rr}}}\bigg|_{\theta=\frac{\pi}{2}},$$

$$u_{in}^\phi = -g^{t\phi}E_{ISCO} + g^{\phi\phi}L_{ISCO}\bigg|_{\theta=\frac{\pi}{2}}, \quad u_{in}^\theta = 0. \quad (3.5)$$

First of all, we notice that the celestial coordinates of each ray have been defined previously. As we track light ray, it's important to note that light can cross the equatorial plane more than once. The radius of each intersection can be written in the form $r_n(x, y)$, $n = 1, 2, 3, ...N_{max}(x, y)$. $r_n(x, y)$ represents the $n-th$ image given on the screen. The image at $n = 1$ is called "direct" image, and the image at $n = 2$ is called "lensed" image.

In general, there are some light rays emitted from the accretion disk can always interacts with the disk. So the intensity would change due to the emission and absorption as the interactions occurs. By ignoring refraction effect of the accretion disk, the relation of the light intensity can be described as the following equation

$$\frac{d}{d\tau}\left(\frac{I_\nu}{\nu^3}\right) = \frac{J_\nu - \kappa_\nu I_\nu}{\nu^2}, \quad (3.6)$$

$I_\nu$ is the specific intensity, $J_\nu$ is the emissivity and $\kappa_\nu$ is the absorption coeffient at the frequency $\nu$. In a vacuum, both $J_\nu$ and $\kappa_\nu$ are zero. This means $I_\nu/\nu^3$ is conserved along the geodesic. It is assumed that the accretion disk axially symmetric, stable, and has $Z_2$-symmetry in the equatorial plane. The accretion disk is geometrically thin, the absorption coefficient and emissivity of accretion disk remain constant as light passes through it. In those considerations, the formula for the light intensity received on the screen is given by

$$I_{\nu_0} = \sum_{n=1}^{N_{max}} \left(\frac{\nu_o}{\nu_n}\right)^3 \frac{J_n}{\iota_{n-1}}\left[\frac{1-e^{\kappa_n f_n}}{\kappa_n}\right], \quad (3.7)$$

which $\nu_o = \mathcal{E}_o = -p_0|_{r=r_o}$ is the observed frequence on the screen, $\nu_n = \mathcal{E}_n = -k_\mu u^\mu|_{r=r_n}$ is the frequence observed by the local rest frames comoving with the accretion disk, $f_n = \nu_n \Delta\lambda_n$ is the "fudge factor" which is related to the accretion disk model. The affine parameters will change as the light passes through the accretion disk at $F_n$, which is $\Delta\lambda_n$. Here, the symbol $F_n$ is called the class of the local rest frames, and $n = 1, 2, 3, ...N_{max}$ is the number of times that the light passes through the accretion disk. And, the optical depth $\iota_m$ is

$$\iota_m = \begin{cases} \exp\left[\sum_{n=1}^m \kappa_n f_n\right] & \text{if } m \geq 1, \\ 1 & \text{if } m = 0. \end{cases} \quad (3.8)$$

When further taking the redshift factor ($g_n = \nu_o/\nu_n$) into account, the Eq. (3.7) can be simplified as $\sum_{n=1}^{N_{max}} f_n g_n^3 J_n$.

In order to compare with the observed data of M87* and Sgr A* images, we choose the specific form of emissivity as

$$J = \exp\left[-\frac{1}{2}\left(\log\frac{r}{r_h}\right)^2 - 2\left(\log\frac{r}{r_h}\right)\right], \quad (3.9)$$

where $r_h$ is the horizon radius on the equatorial plane. The effect of the $f_n$ is mainly to change the light intensity of the narrow photon ring, which has little effect on the overall pattern. According to Ref. [112], the $f_n$ is normalized to 1. In addition, the form of redshift factor for $r_n \geq r_{ISCO}$ is

$$g_n = \frac{e}{\zeta(1-\Omega_n\xi)}, \quad r_n \geq r_{ISCO}, \quad (3.10)$$

and for $r_n < r_{ISCO}$, we have

$$g_n = -\frac{e}{u_{in}^r \frac{k_r}{\mathcal{E}} + E_{ISCO}(g^{tt} - g^{t\phi}\xi) + L_{ISCO}(g^{\phi\phi}\xi - g^{t\phi})},$$
$$r_n < r_{ISCO}. \quad (3.11)$$

with

$$\zeta = \sqrt{\frac{-1}{g_{tt} + 2g_{t\phi}\Omega_n + g_{\phi\phi}\Omega_n^2}}\bigg|_{r=r_n}, \quad (3.12)$$

here the ratio between the energy observed on the screen and the energy along null geodesic is denoted as $e = \mathcal{E}_o/\mathcal{E}$, which is $e = 1$ for KN-MOG spacetime since it is asymptotically flat.

## 4 Accretion disk images

In this paper, we fix the position of the observer at $r_o = 100$. Because the emission profile of the accretion disk decreases rapidly with $r$. So, the accretion disk far from the event horizon has little effect on the image of KN-MOG BH. In view of this, the inner and outer radius of the accretion disk can be set to $r_{ir} = r_h$ and $r_{or} = 20$, which is enough for the purpose of our study.

Figures 4 and 5 are the images of the KN-MOG BH illuminated by an accretion disk with prograde and retrograde flows, respectively. Some photons emitted inside the critical curve are captured by the event horizon, while some photons will escape to infinity and be observed by the observer. As a result, those photons that cannot reach the observer form the black region, known as an inner shadow.

In Fig. 4, for $\theta_o = 80°$, we can easily distinguish the direct image from the lens image. Similarly, the photon ring and the inner shadow can be easily observed. The inner shadow observed in this situation is a blurry black semicircle. The critical curve takes on a "D" shape for $a = 0.998$, but changes

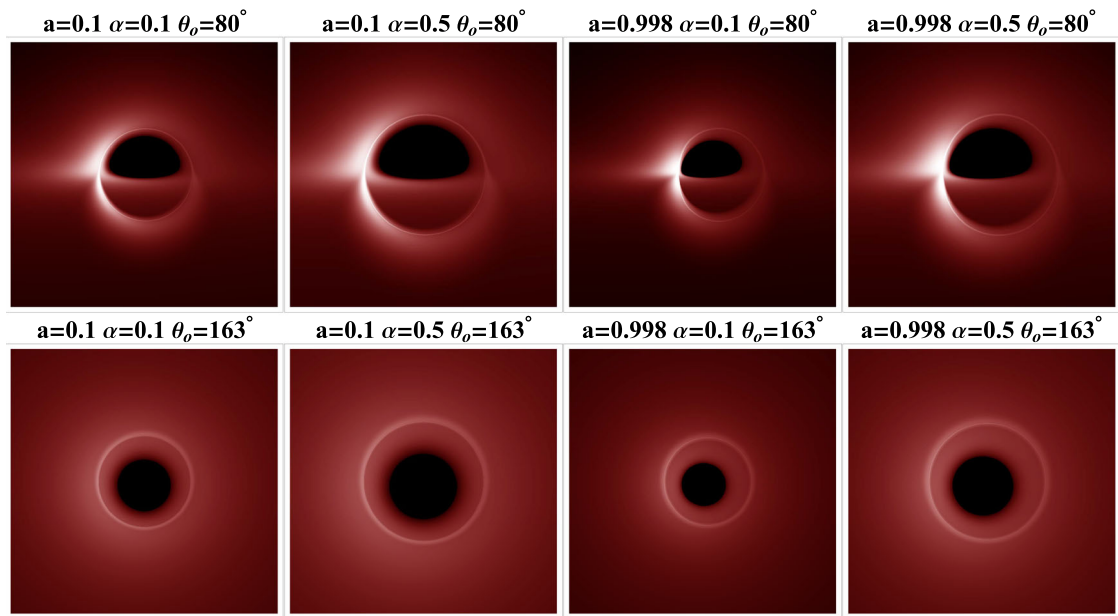

**Fig. 4** Images of KN-MOG BH illuminated by prograde flows with $Q = 0.3Q_e$. The observed positions of the top and bottom rows correspond to $(r_o = 100, \theta_o = 80°)$ and $(r_o = 100, \theta_o = 163°)$, respectively. The first and second columns correspond to the case $(a = 0.1)$. The third and fourth columns correspond to the case $(a = 0.998)$. Meanwhile, the first and third columns correspond to the case $(\alpha = 0.1)$. The second and fourth columns correspond to the case $(\alpha = 0.5)$. The black disks represent the inner shadow of the BH. The bright light ring represents the critical curve. The field of view of camera is 18°

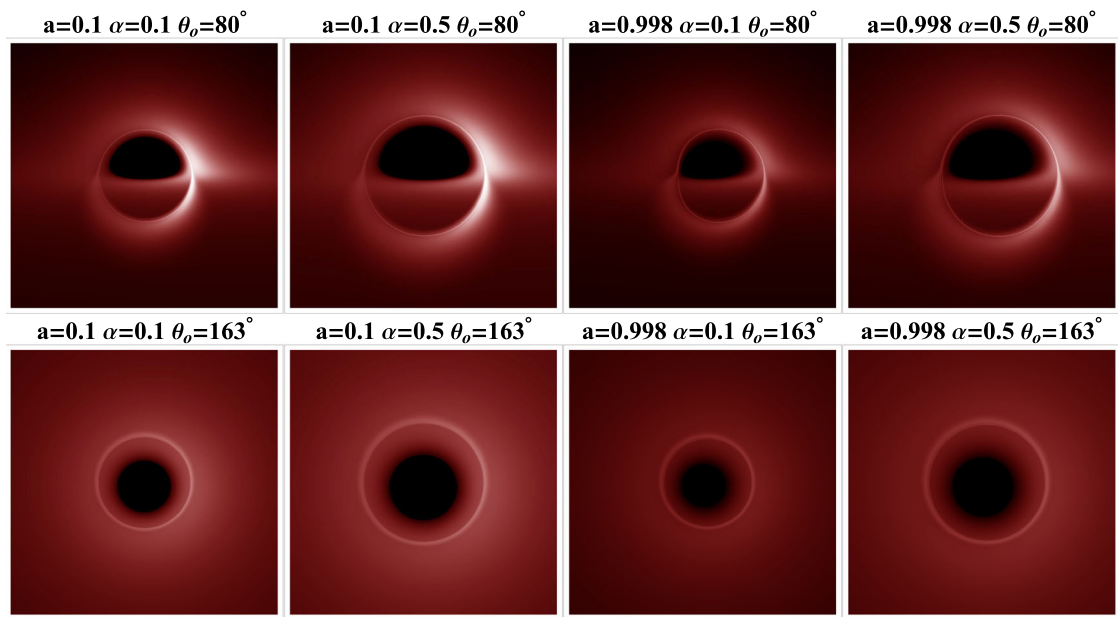

**Fig. 5** Images of KN-MOG BH illuminated by retrograde flows with $Q = 0.3Q_e$. The observed positions of the top and bottom rows correspond to $(r_o = 100, \theta_o = 80°)$ and $(r_o = 100, \theta_o = 163°)$, respectively. The first and second columns correspond to the case $(a = 0.1)$. The third and fourth columns correspond to the case $(a = 0.998)$. Meanwhile, the first and third columns correspond to the case $(\alpha = 0.1)$. The second and fourth columns correspond to the case $(\alpha = 0.5)$. The black disks represent the inner shadow of the BH. The bright light ring represents the critical curve. The field of view of camera is 18°

into a circular shape as $\alpha = 0.5$. As $\alpha$ increases, both the inner shadow and the critical curve become larger and larger. For $\theta_o = 163°$, the direct image and the lens image can no longer be distinguished, only the inner shadow and the photon ring can be observed. The inner shadow observed in this situation is circular. We can also observe that the inner shadow and the critical curve increase as $\alpha$ grows. However, since the change in the shape of the critical curve with $a$ is not observable, the ability of $\alpha$ that restore the circularity is also missing. Figure 5 presents the image of the accretion disk in the retrograde flows. It can be observed that the inner shadow and the critical curve in the accretion disk image change with $a$ and $\alpha$ in the same way as that in the prograde flows. However, in terms of intensity distribution, the prograde and retrograde accretion disks exhibit different behaviors. For the prograde flows, the light intensity observed on the left side of the critical curve is significantly stronger than that on the right side. In contrast, for the retrograde flows, the light intensity observed on the right side is noticeably stronger than that on the left side. This is because the prograde accretion disk rotates to the left, and the left side, due to the Doppler effect, appears to accumulate strength here. Similarly, a retrograde accretion disk rotates to the right, accumulating strength on the right side. However, the strength is somewhat weaker by comparing with the prograde flows due to the drag effect of KN-MOG BH.

The images in Figs. 6 and 7 are the results obtained with $\theta_o = 80°$. In Fig. 6, we show the redshift of the direct images of the accretion disk. It can be seen from the two columns on the left, when $a$ increase, the inner shadow decrease, so the region of redshift seems to be smaller, while an increase $\alpha$ enhance the region of redshift. From the two columns on the right, we can see that both $a$ and $\alpha$ increase the region of the redshift. For the redshift of lensed images in Fig. 7, the results are similar with that of the direct images.

We focus on the intensity distribution in the $y$ direction on the screen, which is shown in Fig. 8. When the accretion disk is in the prograde flows, it can be seen that the increase of $\alpha$ will cause the distance between two peaks to widen. This corresponds to the results that $\alpha$ enlarges the size of BH's shadow. We also find that when $a$ is large, $\alpha$ enhance the peak intensity and the radiative flux. But, when $a$ is small, the peak intensity and the radiative flux decreases with $\alpha$. This implies the observed intensity is not only influenced by $\alpha$ but also related to $a$. When the accretion disk is in the retrograde flows, the effects of $a$ and $\alpha$ on the peak intensity and radiative flux in the same way as in the prograde flows. But, the intensity in the retrograde flows is slightly weaker than in the prograde flows.

To further analyze the effect of charge $Q$ on the images of BH, we show a comparative study of Kerr, KN, Kerr-MOG, and KN-MOG BHs in Fig. 9. We observe that the radius of shadow and inner shadow of BH increases with $\alpha$, while $Q$ decreases them. At the same parameter level, one can see that $\alpha$ has an obvious effect on the shadow and image of BH by comparing with that of $Q$. The bright light ring is called as the critical curve of BH. For a large value of $a$, the left side of the critical curve seems to be flattened for Kerr BH. However, $\alpha$ can gradually make this flattened section become more rounded. This effect is clearer for the case $a = 0.998$. An increase in $Q$ will decrease the critical curve, but the effect is minimal. In view of this, we can conclude that both $\alpha$ and $Q$ have relatively significant effects on the image of the KN-MOG black hole with the thin disk accretion, but the influence of $\alpha$ is clearly much greater. This indicates that $\alpha$ plays a dominant role in this spacetime.

## 5 Discussion and summary

In this paper, we used the backward ray-tracing method to study the shadow and images of accretion disk of KN BH in modified gravity theory. First, the photon motion equation is obtained by solving the Hamilton-Jacobi equation. Then, the BH's shadow is observed with the aid of the ZAMO's frame. we find that the shadow expands as $\alpha$ increases in Fig. 1. Because of the rapidly spin of BHs, the photon trajectories near the BH's equatorial plane deviate from symmetric distribution. So, the left side of shadow gradually flattens as $a$ increases. And, it is interesting to show that this flattening effect gradually diminishes as $\alpha$ increases. From this, it can be inferred that an increase in $\alpha$ enhances the gravitational effect while weakening the dragging effect. Additionally, this effect can be described by the deviation rate $\delta_s$ in Table 1, which increases with $a$ but decreases with $\alpha$. Meanwhile, it also shows that the size of shadow decreases as $Q$ increases in Fig. 2. Also, we studied the BH's images under the astrophysical light source. As $\alpha$ increases, the "D" shaped shadow caused by $a$ diminishes. In Fig. 3, we find a "tail" along the Einstein ring. And, this "tail" becomes longer as $a$ increases, while seems to be unrelated to $\alpha$.

Finally, we further investigated the images of KN-MOG BH with the thin accretion disk. Under both prograde and retrograde flows, the observed images of accretion disk were plotted for different values of $\alpha$, $a$, $Q$ and $\theta_o$. We can easily distinguish the direct image from lens image at $\theta_o = 80°$, which is no longer distinguished at $\theta_o = 163°$. An interesting phenomenon observed from Figs. 4 and 5 is that at larger observed angles, the direct image undergoes a hat-shaped deformation due to the gravitational lensing effect. In Fig. 4, it shows that the light intensity observed on the left side of the critical curve is significantly stronger than that on the right side. This is because the prograde accretion disk rotates to the left, leading to an accumulation of intensity on the left side due to the Doppler effect. In contrast, in Fig. 5, the retrograde accretion disk, which rotates to the right, accumulates the intensity on the right side. However, because of the drag effect of the KN-MOG BH, the intensity on this side is somewhat weaker by compared with that of prograde flows. From Figs. 6 and 7, in the direct images and lensed images of the redshift distribution, it can be observed that for prograde flows, the blueshift appears on the left side of the inner shadow, while for retrograde flows, it appears on the right side of the inner shadow. Additionally, in the case of prograde flows, the red region on the left side of the inner shadow is smaller, whereas for retrograde flows, the red region on the right side of the inner shadow is larger. In addition, it can be concluded that the increase of $\alpha$ enhances the redshift region for both prograde and retrograde flows, while $a$ only increases it in retrograde flows. Overall, the size of the image is determined by the MOG and charge parameters, while the image of KN-MOG BH does not only depend on those parameters, but also related to the position of the observer. Obviously, since the MOG parameter enhances the gravitational effects of the KN-MOG BH, the Doppler effect between the light source and the distant observer is also increased with it, thereby explaining the dependence of the redshift and blueshift distributions on the MOG parameter.

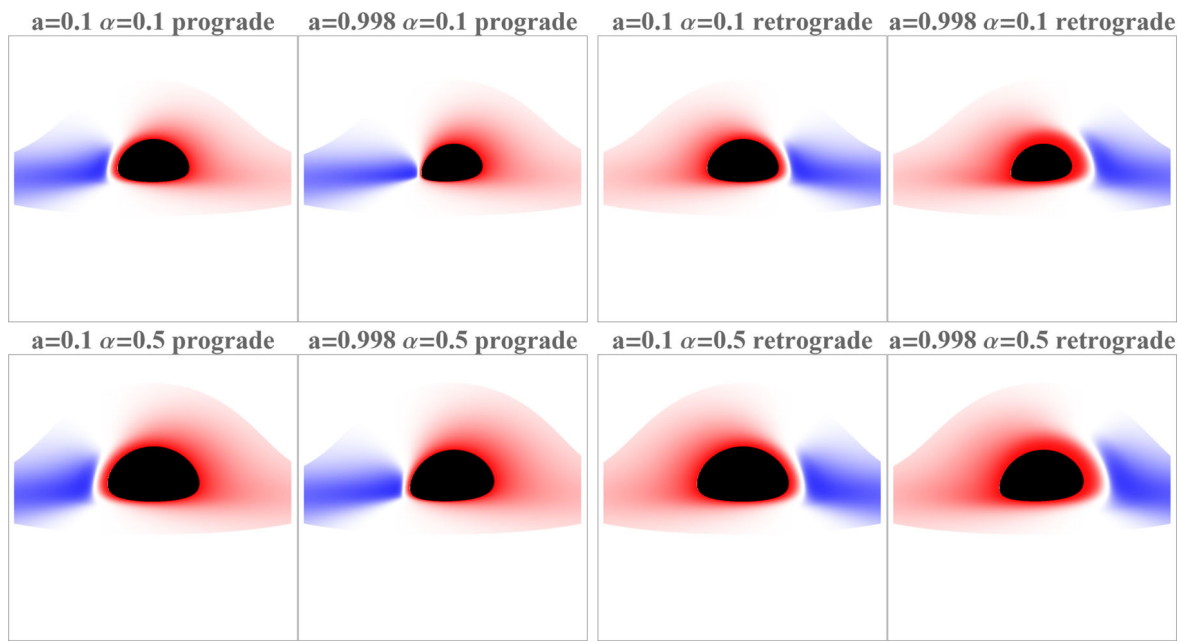

**Fig. 6** The redshift of direct images of the accretion disk with $Q = 0.3Q_e$. The first and second columns correspond to the prograde accretion flows, while the third and fourth columns correspond to the retrograde accretion flows. The top and bottom rows correspond to the case ($\alpha = 0.1$ and $\alpha = 0.5$), respectively. The first and third columns correspond to the case ($a = 0.1$). The second and fourth columns correspond to the case ($a = 0.998$). The red region represents redshift, while the blue region represents blueshift. The position of the observer is fixed at ($r_o = 100$, $\theta_o = 80°$). The camera's field of view is $18°$

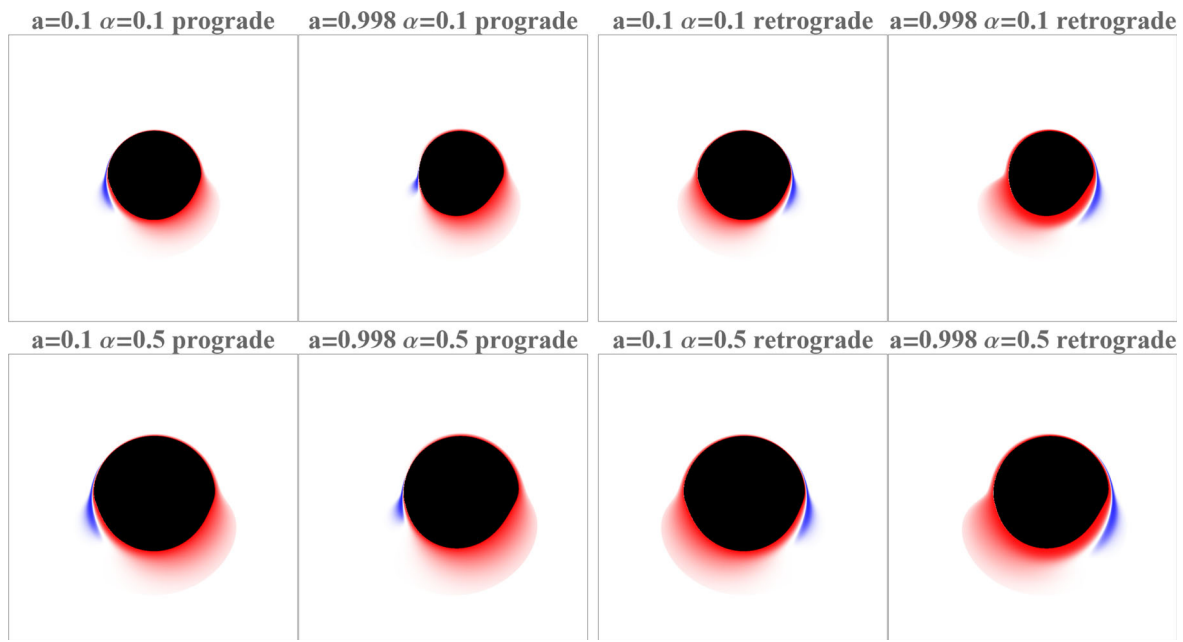

**Fig. 7** The redshift of lensed images of the accretion disk with $Q = 0.3Q_e$. The first and second columns correspond to the prograde accretion flows, while the third and fourth columns correspond to the retrograde accretion flows. The top and bottom rows correspond to the case ($\alpha = 0.1$ and $\alpha = 0.5$), respectively. The first and third columns correspond to the case ($a = 0.1$). The second and fourth columns correspond to the case ($a = 0.998$). The red region represents redshift, while the blue region represents blueshift. The position of the observer is fixed at ($r_o = 100$, $\theta_o = 80°$). The camera's field of view is $18°$

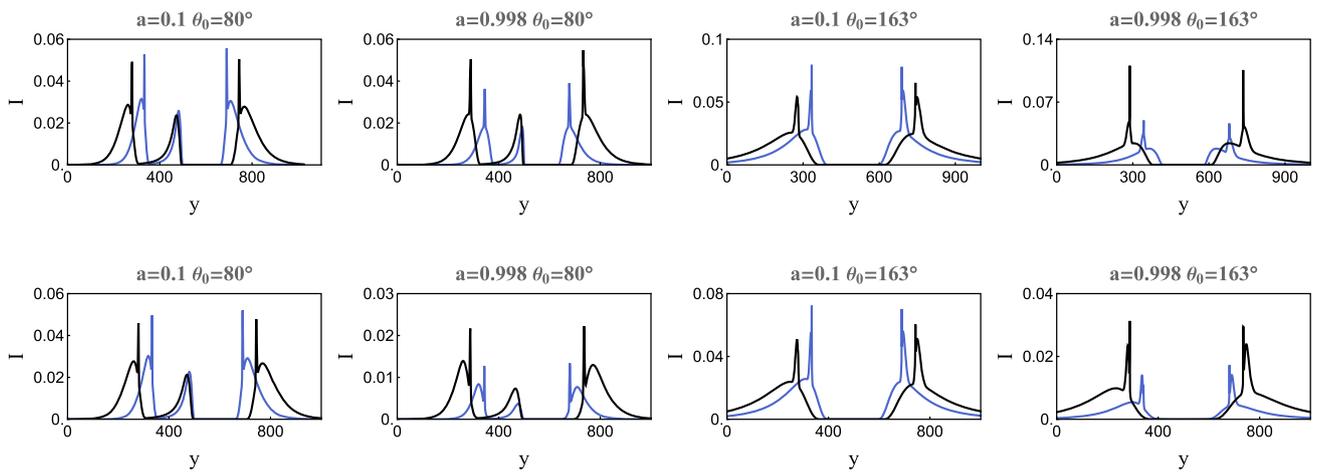

**Fig. 8** Intensity distribution along $y$-axis of the screen with $Q = 0.3Q_e$. The position of observer for the images in the left two columns and the right two columns are at $(r_o = 100, \theta_o = 80°)$ and $(r_o = 100, \theta_o = 163°)$, respectively. The camera's field of view is $18°$. The top and bottom rows correspond to prograde and retrograde flows respectively, where $\alpha = 0.1$ (solid blue line), $\alpha = 0.5$ (solid black line). The first and third columns correspond to the case ($a = 0.1$), while the second and fourth columns correspond to the case ($a = 0.998$)

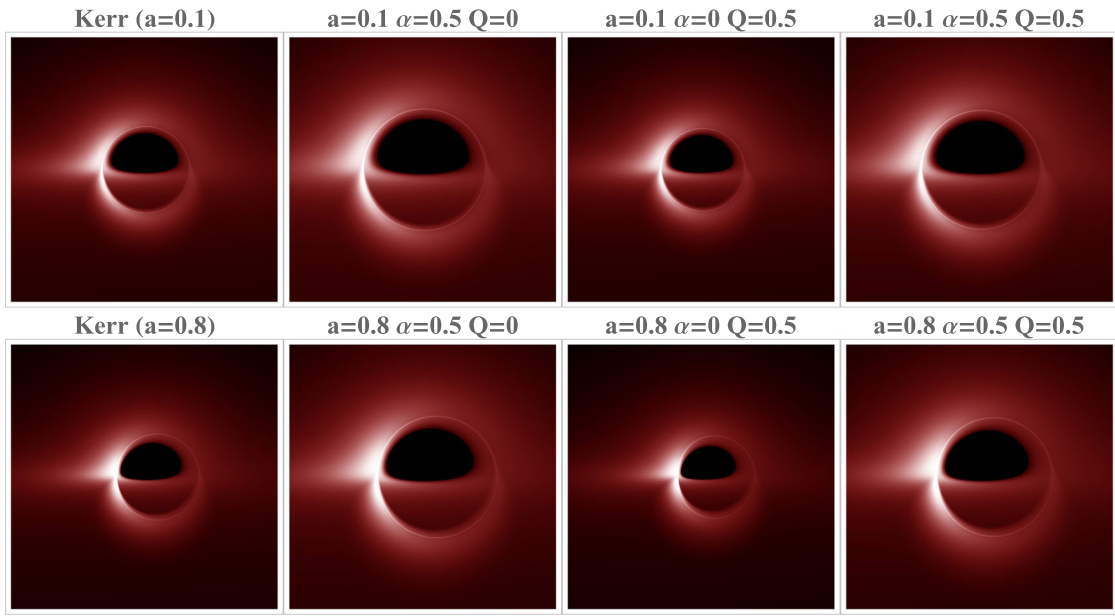

**Fig. 9** Images of BHs illuminated by prograde flows. The top row, from left to right, consists of: Kerr BH ($a = 0.1$), Kerr-MOG BH ($a = 0.1, \alpha = 0.5$), KN BH ($a = 0.1, Q = 0.5$), and KN-MOG BH ($a = 0.1, \alpha = 0.5, Q = 0.5$). The bottom row, from left to right, consists of: Kerr BH ($a = 0.8$), Kerr-MOG BH ($a = 0.8, \alpha = 0.5$), KN BH ($a = 0.8, Q = 0.5$), and KN-MOG BH ($a = 0.8, \alpha = 0.5, Q = 0.5$). The position of the observer is fixed at $(r_o = 100, \theta_o = 80°)$. The camera's field of view is $18°$

In Fig. 8, for the intensity distribution in the $y$ direction on the screen, the increase of $\alpha$ cause the distance between two peaks to widen. And when $a$ is large, $\alpha$ enhance the peak intensity and the radiative flux. But, when $a$ is small, the peak intensity and the radiative flux decreases with $\alpha$. Moreover, we observe that the radius of shadow and inner shadow of BH increases with $\alpha$, but decreases with $Q$. In Fig. 9, by comparing the accretion disks of Kerr BHs, Kerr-MOG BHs, KN BHs, and KN-MOG BHs, we found that Kerr-MOG BHs exhibit the largest inner shadow. In contrast, KN BHs show the smallest inner shadow, while the size of the inner shadow for KN-MOG BHs lies between that of Kerr BHs and Kerr-MOG BHs. At the same parameter level, one can see that $\alpha$ has a more obvious effect on the shadow and image of BH by comparing with that of $Q$. For a large value of $a$, the left side of the critical curve seems to be flattened for Kerr BH. The parameter $\alpha$ can gradually make this flattened section become more rounded. An increase in $Q$ decreases the critical curve, but the effect is minimal. In view of this, we can conclude that both $\alpha$ and $Q$ have relatively significant effects on the image of the KN-MOG BH with the thin disk accretion, but the influence of $\alpha$ is clearly much greater. This indicates that $\alpha$ plays a dominant role in this spacetime, which can also be seen from the metric equations (2.1) and (2.2). In conclusion, the key finding of this paper is that the MOG parameter $\alpha$ has a significant effect on the image of the KN-MOG BH, including the shadow and light rings. These differences provide potential observational features to distinguish Kerr-MOG BHs from BHs in standard General Relativity and serve as an effective tool for probing the feature of MOG-modified gravity. These effects provide a new theoretical support for astronomical observations. In the future, it is also interesting to further study the image of the KN-MOG BH with thick accretion disks surrounding it, which may also present some different observed features of MOG-modified gravity.

**Acknowledgements** This work was supported by the National Natural Science Foundation of China (Grant no. 11903025) and the Sichuan Science and Technology Program (Grant no. 2023ZYD0023).

**Data Availability Statement** Data will be made available on reasonable request. [Author's comment: The datasets generated during and/or analysed during the current study are available from the corresponding author on reasonable request.]

**Code Availability Statement** Code/software will be made available on reasonable request. [Author's comment: The code/software generated during and/or analysed during the current study is available from the corresponding author on reasonable request.]